\documentclass{article}[10pt,letterpaper]
\usepackage{fancyhdr}
\usepackage{supertabular}
\usepackage{color}
\usepackage{amsmath}
\usepackage{pbox}
\usepackage{amssymb}
\usepackage{amsbsy}
\usepackage{scalefnt}
\usepackage{ifpdf}
\usepackage{graphicx}
\usepackage{multicol}
\usepackage[colorlinks=true, linkcolor=blue, linktocpage=true,urlcolor=blue]{hyperref}
\usepackage{longtable}
\usepackage{needspace}
\usepackage{eso-pic}
\usepackage{mathptmx}
\usepackage{multirow}
\usepackage{lscape}
\usepackage{textcomp}
\usepackage{changepage}
\usepackage{xfrac}
\usepackage{txfonts}
\usepackage{pdfcomment}
\usepackage[absolute]{textpos}
\ifpdf
\fi
\makeatletter
\renewcommand{\l@section}{\@dottedtocline{2}{1em}{0em}}
\renewcommand{\l@subsection}{\@dottedtocline{3}{2.5cm}{0em}}
\renewcommand{\hrulefill}{\leavevmode \leaders \hrule \@height 1pt \hfill \kern\z@}
\makeatother
\renewcommand{\underline}[1]{\begin{tabular}{@{\extracolsep{\fill}}c@{\extracolsep{\fill}}}#1\\[-0.2cm]\hrulefill\end{tabular}}

\newcounter{chappage}
\AddToShipoutPicture{\stepcounter{chappage}}
\fancypagestyle{plain}{\lhead{}\rhead{}}
\fancypagestyle{single}{\lhead{\leftmark \ \\}\rhead{\leftmark \ \\}}
\fancypagestyle{bob}{\lhead{\leftmark -\thechappage \ \\}\rhead{\leftmark -\thechappage \ \\}}
\begin{document}
\fontsize{9}{10}\selectfont
\fontdimen2\font=1.3\fontdimen2\font
%Abstract
\thispagestyle{empty}
\setcounter{page}{1}
\begin{center}

\vspace{0.5cm}
{ \huge Energy Levels of \ensuremath{^{\textnormal{8}}}C*}\\
\vspace{1.0cm}
{ \normalsize K. Setoodehnia\ensuremath{^{\textnormal{1,2}}} and J. H. Kelley\ensuremath{^{\textnormal{1,3}}}}\\
\vspace{0.2in}
{ \small \it \ensuremath{^{\textnormal{1}}}Triangle Universities Nuclear Laboratory, Duke University,\\
  Durham North Carolina 27708, USA.\\
  \ensuremath{^{\textnormal{2}}}Department of Physics, Duke University, Durham, North Carolina\\
  27708, USA.\\
  \ensuremath{^{\textnormal{3}}}Department of Physics, North Carolina State University,\\
  Raleigh, North Carolina 27607, USA}\\
\vspace{0.2in}
\end{center}

\setlength{\parindent}{-0.5cm}
\addtolength{\leftskip}{2cm}
\addtolength{\rightskip}{2cm}
{\bf Abstract: }
In this document, experimental nuclear structure data are evaluated for \ensuremath{^{\textnormal{8}}}C. The details of each reaction populating \ensuremath{^{\textnormal{8}}}C levels are compiled and evaluated. The combined results provide a set of adopted values that include level energies, spins and parities, level widths, decay types and branching ratios, and other nuclear properties. This work supersedes the earlier work by J. E. Purcell in (2018) published in ENSDF database.\\

{\bf Cutoff Date: }
Literature available up to October 31, 2024 has been considered; the primary bibliographic source, the NSR database (\href{https://www.nndc.bnl.gov/nsr/nsrlink.jsp?2011Pr03,B}{2011Pr03}) available at Brookhaven National Laboratory web page: www.nndc.bnl.gov/nsr/.\\

{\bf General Policies and Organization of Material: }
See the April 2025 issue of the {\it Nuclear Data Sheets} or \\https://www.nndc.bnl.gov/nds/docs/NDSPolicies.pdf. \\

{\bf Acknowledgements: }
The authors expresses her gratitude to personnel at the National Nuclear Data Center (NNDC) at Brookhaven National Laboratory for facilitating this work.\\

\vfill

* This work is supported by the Office of Nuclear Physics, Office of Science, U.S. Department of Energy under contracts: DE-FG02-97ER41042 {\textminus} North Carolina State University and DE-FG02-97ER41033 {\textminus} Duke University\\

\setlength{\parindent}{+0.5cm}
\addtolength{\leftskip}{-2cm}
\addtolength{\rightskip}{-2cm}
\newpage
\pagestyle{plain}
\setlength{\columnseprule}{1pt}
\setlength{\columnsep}{1cm}
\begin{center}
\underline{\normalsize Index for A=8}
\end{center}
\hspace{.3cm}\raggedright\underline{Nuclide}\hspace{1cm}\underline{Data Type\mbox{\hspace{2.3cm}}}\hspace{2cm}\underline{Page}\hspace{1cm}
\raggedright\underline{Nuclide}\hspace{1cm}\underline{Data Type\mbox{\hspace{2.3cm}}}\hspace{2cm}\underline{Page}
\begin{adjustwidth}{}{0.05\textwidth}
\begin{multicols}{2}
\setcounter{tocdepth}{3}
\renewcommand{\contentsname}{\protect\vspace{-0.8cm}}
\tableofcontents
\end{multicols}
\end{adjustwidth}
\clearpage
\thispagestyle{empty}
\mbox{}
\clearpage
\clearpage
\pagestyle{bob}
\begin{center}
%ADOPTED LEVELS
\section[\ensuremath{^{8}_{6}}C\ensuremath{_{2}^{~}}]{ }
\vspace{-30pt}
\setcounter{chappage}{1}
\subsection[\hspace{-0.2cm}Adopted Levels]{ }
\vspace{-20pt}
\vspace{0.3cm}
\hypertarget{C0}{{\bf \small \underline{Adopted \hyperlink{8C_LEVEL}{Levels}}}}\\
\vspace{4pt}
\vspace{8pt}
\parbox[b][0.3cm]{17.7cm}{\addtolength{\parindent}{-0.2in}S(p)=$-$100 {\it 30}\hspace{0.2in}\href{https://www.nndc.bnl.gov/nsr/nsrlink.jsp?2021Wa16,B}{2021Wa16}}\\
\parbox[b][0.3cm]{17.7cm}{\addtolength{\parindent}{-0.2in}S(2p)={\textminus}2111 keV \textit{19} (\href{https://www.nndc.bnl.gov/nsr/nsrlink.jsp?2021Wa16,B}{2021Wa16}).}\\

\parbox[b][0.3cm]{17.7cm}{\addtolength{\parindent}{-0.2in}The atomic mass excess of \ensuremath{^{\textnormal{8}}}C is given as 35064 keV \textit{18} (\href{https://www.nndc.bnl.gov/nsr/nsrlink.jsp?2021Wa16,B}{2021Wa16}) from the average of 35030 keV \textit{30} (\href{https://www.nndc.bnl.gov/nsr/nsrlink.jsp?2011Ch32,B}{2011Ch32}), 35.10 MeV}\\
\parbox[b][0.3cm]{17.7cm}{\textit{3} (\href{https://www.nndc.bnl.gov/nsr/nsrlink.jsp?1976Tr01,B}{1976Tr01}), and 35.06 MeV \textit{5} (\href{https://www.nndc.bnl.gov/nsr/nsrlink.jsp?1976Ro04,B}{1976Ro04}). The mass excess of \ensuremath{^{\textnormal{8}}}C was the same in (\href{https://www.nndc.bnl.gov/nsr/nsrlink.jsp?2017Wa10,B}{2017Wa10}). Using this mass excess value,}\\
\parbox[b][0.3cm]{17.7cm}{the binding energy of \ensuremath{^{\textnormal{8}}}C is 24.812 MeV \textit{18} (\href{https://www.nndc.bnl.gov/nsr/nsrlink.jsp?2021Wa16,B}{2021Wa16}). In some of the theoretical articles referenced below, this is the quantity}\\
\parbox[b][0.3cm]{17.7cm}{that is calculated. Sometimes the binding energy is given relative to the \ensuremath{^{\textnormal{4}}}He+4p threshold. The latter is 3.483 MeV \textit{18} based on}\\
\parbox[b][0.3cm]{17.7cm}{AME-2020 (\href{https://www.nndc.bnl.gov/nsr/nsrlink.jsp?2021Wa16,B}{2021Wa16}).}\\
\parbox[b][0.3cm]{17.7cm}{\addtolength{\parindent}{-0.2in}\ensuremath{^{\textnormal{8}}}C is unstable with respect to single proton decay (Q=100 keV \textit{30}), two proton decay (Q=2111 keV \textit{19}), three proton decay}\\
\parbox[b][0.3cm]{17.7cm}{(Q=1518 keV \textit{53}), and four proton decay (Q=3483 keV \textit{18}). The Q-values that are mentioned here are from AME-2020}\\
\parbox[b][0.3cm]{17.7cm}{(\href{https://www.nndc.bnl.gov/nsr/nsrlink.jsp?2021Wa16,B}{2021Wa16}). Results reported in (\href{https://www.nndc.bnl.gov/nsr/nsrlink.jsp?2010Ch42,B}{2010Ch42}, \href{https://www.nndc.bnl.gov/nsr/nsrlink.jsp?2023Ch46,B}{2023Ch46}) indicate that \ensuremath{^{\textnormal{8}}}C decays by emitting two pairs of protons {\textminus}}\\
\parbox[b][0.3cm]{17.7cm}{\ensuremath{^{\textnormal{8}}}C\ensuremath{\rightarrow}\ensuremath{^{\textnormal{6}}}Be+2p\ensuremath{\rightarrow}(\ensuremath{^{\textnormal{4}}}He+2p)+2p. Also see (\href{https://www.nndc.bnl.gov/nsr/nsrlink.jsp?2011ChZW,B}{2011ChZW}).}\\
\parbox[b][0.3cm]{17.7cm}{\addtolength{\parindent}{-0.2in}For theoretical studies that include \ensuremath{^{\textnormal{8}}}C see (\href{https://www.nndc.bnl.gov/nsr/nsrlink.jsp?1974Ce05,B}{1974Ce05}, \href{https://www.nndc.bnl.gov/nsr/nsrlink.jsp?1974Ir04,B}{1974Ir04}, \href{https://www.nndc.bnl.gov/nsr/nsrlink.jsp?1987Bl18,B}{1987Bl18}, \href{https://www.nndc.bnl.gov/nsr/nsrlink.jsp?1987Sa15,B}{1987Sa15}, \href{https://www.nndc.bnl.gov/nsr/nsrlink.jsp?1988Co15,B}{1988Co15}, \href{https://www.nndc.bnl.gov/nsr/nsrlink.jsp?1996Gr21,B}{1996Gr21}, \href{https://www.nndc.bnl.gov/nsr/nsrlink.jsp?1996Ka14,B}{1996Ka14}, \href{https://www.nndc.bnl.gov/nsr/nsrlink.jsp?1996Su24,B}{1996Su24},}\\
\parbox[b][0.3cm]{17.7cm}{\href{https://www.nndc.bnl.gov/nsr/nsrlink.jsp?1997Ba54,B}{1997Ba54}, \href{https://www.nndc.bnl.gov/nsr/nsrlink.jsp?1997Po12,B}{1997Po12}, \href{https://www.nndc.bnl.gov/nsr/nsrlink.jsp?1998Wi10,B}{1998Wi10}, \href{https://www.nndc.bnl.gov/nsr/nsrlink.jsp?1999Ha61,B}{1999Ha61}, \href{https://www.nndc.bnl.gov/nsr/nsrlink.jsp?2000Wi09,B}{2000Wi09}, \href{https://www.nndc.bnl.gov/nsr/nsrlink.jsp?2001Co21,B}{2001Co21}, \href{https://www.nndc.bnl.gov/nsr/nsrlink.jsp?2002Ba90,B}{2002Ba90}, \href{https://www.nndc.bnl.gov/nsr/nsrlink.jsp?2002Fo11,B}{2002Fo11}, \href{https://www.nndc.bnl.gov/nsr/nsrlink.jsp?2003Ba99,B}{2003Ba99}, \href{https://www.nndc.bnl.gov/nsr/nsrlink.jsp?2006Sa29,B}{2006Sa29}, \href{https://www.nndc.bnl.gov/nsr/nsrlink.jsp?2006Wi07,B}{2006Wi07},}\\
\parbox[b][0.3cm]{17.7cm}{\href{https://www.nndc.bnl.gov/nsr/nsrlink.jsp?2007Ma79,B}{2007Ma79}, \href{https://www.nndc.bnl.gov/nsr/nsrlink.jsp?2009Ba41,B}{2009Ba41}, \href{https://www.nndc.bnl.gov/nsr/nsrlink.jsp?2010Ti04,B}{2010Ti04}, \href{https://www.nndc.bnl.gov/nsr/nsrlink.jsp?2011ChZW,B}{2011ChZW}, \href{https://www.nndc.bnl.gov/nsr/nsrlink.jsp?2012My02,B}{2012My02}, \href{https://www.nndc.bnl.gov/nsr/nsrlink.jsp?2012My04,B}{2012My04}, \href{https://www.nndc.bnl.gov/nsr/nsrlink.jsp?2014Eb02,B}{2014Eb02}, \href{https://www.nndc.bnl.gov/nsr/nsrlink.jsp?2014Mi17,B}{2014Mi17}, \href{https://www.nndc.bnl.gov/nsr/nsrlink.jsp?2014My03,B}{2014My03}, \href{https://www.nndc.bnl.gov/nsr/nsrlink.jsp?2019Ka50,B}{2019Ka50}, \href{https://www.nndc.bnl.gov/nsr/nsrlink.jsp?2019Sh36,B}{2019Sh36},}\\
\parbox[b][0.3cm]{17.7cm}{\href{https://www.nndc.bnl.gov/nsr/nsrlink.jsp?2021My01,B}{2021My01}, \href{https://www.nndc.bnl.gov/nsr/nsrlink.jsp?2021Wy01,B}{2021Wy01}, \href{https://www.nndc.bnl.gov/nsr/nsrlink.jsp?2021Xi06,B}{2021Xi06}, \href{https://www.nndc.bnl.gov/nsr/nsrlink.jsp?2022De06,B}{2022De06}, \href{https://www.nndc.bnl.gov/nsr/nsrlink.jsp?2024Ya25,B}{2024Ya25}). IMME studies including A=8 are reported in (\href{https://www.nndc.bnl.gov/nsr/nsrlink.jsp?1974Ro17,B}{1974Ro17}, \href{https://www.nndc.bnl.gov/nsr/nsrlink.jsp?1976Tr01,B}{1976Tr01},}\\
\parbox[b][0.3cm]{17.7cm}{\href{https://www.nndc.bnl.gov/nsr/nsrlink.jsp?1984An18,B}{1984An18}, \href{https://www.nndc.bnl.gov/nsr/nsrlink.jsp?1998Br09,B}{1998Br09}, \href{https://www.nndc.bnl.gov/nsr/nsrlink.jsp?2011Ch53,B}{2011Ch53}, \href{https://www.nndc.bnl.gov/nsr/nsrlink.jsp?2013La29,B}{2013La29}, \href{https://www.nndc.bnl.gov/nsr/nsrlink.jsp?2022Zo01,B}{2022Zo01}). Calculations of the \ensuremath{^{\textnormal{8}}}C rms radii are reported in (\href{https://www.nndc.bnl.gov/nsr/nsrlink.jsp?2017Ka45,B}{2017Ka45}, \href{https://www.nndc.bnl.gov/nsr/nsrlink.jsp?2023My01,B}{2023My01},}\\
\parbox[b][0.3cm]{17.7cm}{\href{https://www.nndc.bnl.gov/nsr/nsrlink.jsp?2023My02,B}{2023My02}).}\\
\vspace{12pt}
\hypertarget{8C_LEVEL}{\underline{$^{8}$C Levels}}\\
\begin{longtable}[c]{llll}
\multicolumn{4}{c}{\underline{Cross Reference (XREF) Flags}}\\
 \\
\hyperlink{C1}{\texttt{A }}& \ensuremath{^{\textnormal{1}}}H(\ensuremath{^{\textnormal{9}}}C,d) & \hyperlink{C4}{\texttt{D }}& \ensuremath{^{\textnormal{12}}}C(\ensuremath{\alpha},\ensuremath{^{\textnormal{8}}}He)\\
\hyperlink{C2}{\texttt{B }}& \ensuremath{^{\textnormal{9}}}Be(\ensuremath{^{\textnormal{9}}}C,\ensuremath{^{\textnormal{8}}}C) & \hyperlink{C5}{\texttt{E }}& \ensuremath{^{\textnormal{14}}}N(\ensuremath{^{\textnormal{3}}}He,\ensuremath{^{\textnormal{9}}}Li)\\
\hyperlink{C3}{\texttt{C }}& \ensuremath{^{\textnormal{9}}}Be(\ensuremath{^{\textnormal{13}}}O,\ensuremath{^{\textnormal{8}}}C) & \\
\end{longtable}
\vspace{-0.5cm}
\begin{longtable}{ccccccc@{\extracolsep{\fill}}c}
\multicolumn{2}{c}{E(level)$^{}$}&J$^{\pi}$$^{{\hyperlink{C0LEVEL1}{b}}}$&\multicolumn{2}{c}{\ensuremath{\Gamma}$^{}$}&XREF&Comments&\\[-.2cm]
\multicolumn{2}{c}{\hrulefill}&\hrulefill&\multicolumn{2}{c}{\hrulefill}&\hrulefill&\hrulefill&
\endfirsthead
\multicolumn{1}{r@{}}{0}&\multicolumn{1}{@{}l}{}&\multicolumn{1}{l}{0\ensuremath{^{+}}}&\multicolumn{1}{r@{}}{130}&\multicolumn{1}{@{ }l}{keV {\it 50}}&\multicolumn{1}{l}{\texttt{\hyperlink{C1}{A}\hyperlink{C2}{B}\hyperlink{C3}{C}\hyperlink{C4}{D}\hyperlink{C5}{E}} }&\parbox[t][0.3cm]{10.62374cm}{\raggedright \%2p=100\vspace{0.1cm}}&\\
&&&&&&\parbox[t][0.3cm]{10.62374cm}{\raggedright T=2\vspace{0.1cm}}&\\
&&&&&&\parbox[t][0.3cm]{10.62374cm}{\raggedright \ensuremath{\Gamma}: From \ensuremath{^{\textnormal{9}}}Be(\ensuremath{^{\textnormal{9}}}C,\ensuremath{^{\textnormal{8}}}C) (\href{https://www.nndc.bnl.gov/nsr/nsrlink.jsp?2011Ch32,B}{2011Ch32}); other values from \ensuremath{^{\textnormal{12}}}C(\ensuremath{\alpha},\ensuremath{^{\textnormal{8}}}He) \ensuremath{\Gamma}=0.22\vspace{0.1cm}}&\\
&&&&&&\parbox[t][0.3cm]{10.62374cm}{\raggedright {\ }{\ }{\ }MeV \textit{+8{\textminus}14} (\href{https://www.nndc.bnl.gov/nsr/nsrlink.jsp?1974Ro17,B}{1974Ro17}), \ensuremath{^{\textnormal{14}}}N(\ensuremath{^{\textnormal{3}}}He,\ensuremath{^{\textnormal{9}}}Li) \ensuremath{\Gamma}=290 keV \textit{80} (\href{https://www.nndc.bnl.gov/nsr/nsrlink.jsp?1976Ro04,B}{1976Ro04}),\vspace{0.1cm}}&\\
&&&&&&\parbox[t][0.3cm]{10.62374cm}{\raggedright {\ }{\ }{\ }\ensuremath{^{\textnormal{12}}}C(\ensuremath{\alpha},\ensuremath{^{\textnormal{8}}}He): either \ensuremath{\Gamma}=230 keV \textit{50} from a Gaussian shaped fit or \ensuremath{\Gamma}=183\vspace{0.1cm}}&\\
&&&&&&\parbox[t][0.3cm]{10.62374cm}{\raggedright {\ }{\ }{\ }keV \textit{56} from a Breit-Wigner shaped fit (\href{https://www.nndc.bnl.gov/nsr/nsrlink.jsp?1976Tr01,B}{1976Tr01}), and \ensuremath{^{\textnormal{9}}}Be(\ensuremath{^{\textnormal{13}}}O,\ensuremath{^{\textnormal{8}}}C)\vspace{0.1cm}}&\\
&&&&&&\parbox[t][0.3cm]{10.62374cm}{\raggedright {\ }{\ }{\ }\ensuremath{\Gamma}=88 keV \textit{61} (\href{https://www.nndc.bnl.gov/nsr/nsrlink.jsp?2023Ch46,B}{2023Ch46}: supplemental material). The higher statistics in\vspace{0.1cm}}&\\
&&&&&&\parbox[t][0.3cm]{10.62374cm}{\raggedright {\ }{\ }{\ }(\href{https://www.nndc.bnl.gov/nsr/nsrlink.jsp?2011Ch32,B}{2011Ch32}) compared to the earlier results leads to the choice of \ensuremath{\Gamma}=130\vspace{0.1cm}}&\\
&&&&&&\parbox[t][0.3cm]{10.62374cm}{\raggedright {\ }{\ }{\ }keV \textit{50}. The value from \ensuremath{^{\textnormal{9}}}Be(\ensuremath{^{\textnormal{13}}}O,\ensuremath{^{\textnormal{8}}}C) given in the supplemental material of\vspace{0.1cm}}&\\
&&&&&&\parbox[t][0.3cm]{10.62374cm}{\raggedright {\ }{\ }{\ }(\href{https://www.nndc.bnl.gov/nsr/nsrlink.jsp?2023Ch46,B}{2023Ch46}) is neglected here.\vspace{0.1cm}}&\\
&&&&&&\parbox[t][0.3cm]{10.62374cm}{\raggedright The decay proceeds by \ensuremath{^{\textnormal{8}}}C\ensuremath{\rightarrow}2p+\ensuremath{^{\textnormal{6}}}Be\ensuremath{_{\textnormal{g.s.}}}\ensuremath{\rightarrow}4p+\ensuremath{^{\textnormal{4}}}He (\href{https://www.nndc.bnl.gov/nsr/nsrlink.jsp?2010Ch42,B}{2010Ch42}, \href{https://www.nndc.bnl.gov/nsr/nsrlink.jsp?2023Ch46,B}{2023Ch46}).\vspace{0.1cm}}&\\
&&&&&&\parbox[t][0.3cm]{10.62374cm}{\raggedright {\ }{\ }{\ }The latter study deduced E\ensuremath{_{\textnormal{c.m.}}}(4p+\ensuremath{\alpha})=3490 keV \textit{20}, from which we\vspace{0.1cm}}&\\
&&&&&&\parbox[t][0.3cm]{10.62374cm}{\raggedright {\ }{\ }{\ }determined the mass of \ensuremath{^{\textnormal{8}}}C to be 8.03765 u \textit{16}.\vspace{0.1cm}}&\\
\multicolumn{1}{r@{}}{3.40\ensuremath{\times10^{3}}}&\multicolumn{1}{@{}l}{\ensuremath{^{{\hyperlink{C0LEVEL0}{a}}}} {\it 25}}&\multicolumn{1}{l}{2\ensuremath{^{+}}}&\multicolumn{1}{r@{}}{3}&\multicolumn{1}{@{.}l}{0\ensuremath{^{{\hyperlink{C0LEVEL0}{a}}}} MeV {\it 5}}&\multicolumn{1}{l}{\texttt{\hyperlink{C1}{A}\ \ \ \ } }&\parbox[t][0.3cm]{10.62374cm}{\raggedright T=2 (\href{https://www.nndc.bnl.gov/nsr/nsrlink.jsp?2024Ko04,B}{2024Ko04})\vspace{0.1cm}}&\\
&&&&&&\parbox[t][0.3cm]{10.62374cm}{\raggedright J\ensuremath{^{\pi}}: The L=1 neutron transfer deduced by the DWBA analysis of (\href{https://www.nndc.bnl.gov/nsr/nsrlink.jsp?2024Ko04,B}{2024Ko04}:\vspace{0.1cm}}&\\
&&&&&&\parbox[t][0.3cm]{10.62374cm}{\raggedright {\ }{\ }{\ }\ensuremath{^{\textnormal{1}}}H(\ensuremath{^{\textnormal{9}}}C,d)) leads to J\ensuremath{^{\ensuremath{\pi}}}=(0, 1, 2)\ensuremath{^{\textnormal{+}}}. (\href{https://www.nndc.bnl.gov/nsr/nsrlink.jsp?2024Ko04,B}{2024Ko04}) assigned J\ensuremath{^{\ensuremath{\pi}}}=2\ensuremath{^{\textnormal{+}}} based on\vspace{0.1cm}}&\\
&&&&&&\parbox[t][0.3cm]{10.62374cm}{\raggedright {\ }{\ }{\ }comparisons of the deduced experimental spectroscopic factors for each\vspace{0.1cm}}&\\
&&&&&&\parbox[t][0.3cm]{10.62374cm}{\raggedright {\ }{\ }{\ }allowed J\ensuremath{^{\ensuremath{\pi}}} value and those obtained using shell model calculations. Only\vspace{0.1cm}}&\\
&&&&&&\parbox[t][0.3cm]{10.62374cm}{\raggedright {\ }{\ }{\ }for the case of J\ensuremath{^{\ensuremath{\pi}}}=2\ensuremath{^{\textnormal{+}}}, the theoretical spectroscopic factors reproduced the\vspace{0.1cm}}&\\
&&&&&&\parbox[t][0.3cm]{10.62374cm}{\raggedright {\ }{\ }{\ }experimental ones. The other possible assignments would have a negligibly\vspace{0.1cm}}&\\
&&&&&&\parbox[t][0.3cm]{10.62374cm}{\raggedright {\ }{\ }{\ }small spectroscopic factors, and thus were rejected.\vspace{0.1cm}}&\\
&&&&&&\parbox[t][0.3cm]{10.62374cm}{\raggedright T: From comparison of the E\ensuremath{_{\textnormal{x}}}(\ensuremath{^{\textnormal{8}}}C*(2\ensuremath{^{\textnormal{+}}_{\textnormal{1}}})) with those of other proton- and\vspace{0.1cm}}&\\
&&&&&&\parbox[t][0.3cm]{10.62374cm}{\raggedright {\ }{\ }{\ }neutron-rich nuclei (\href{https://www.nndc.bnl.gov/nsr/nsrlink.jsp?2024Ko04,B}{2024Ko04}). The extracted T=2 suggests the closure of\vspace{0.1cm}}&\\
&&&&&&\parbox[t][0.3cm]{10.62374cm}{\raggedright {\ }{\ }{\ }the semi-magic Z=6 sub-shell for this state.\vspace{0.1cm}}&\\
&&&&&&\parbox[t][0.3cm]{10.62374cm}{\raggedright The mirror energy difference, \ensuremath{\Delta}E\ensuremath{_{\textnormal{x}}}=E\ensuremath{_{\textnormal{x}}}(\ensuremath{^{\textnormal{8}}}C*(2\ensuremath{^{\textnormal{+}}_{\textnormal{1}}})){\textminus}E\ensuremath{_{\textnormal{x}}}(\ensuremath{^{\textnormal{8}}}He*(2\ensuremath{^{\textnormal{+}}_{\textnormal{1}}}))={\textminus}0.14\vspace{0.1cm}}&\\
&&&&&&\parbox[t][0.3cm]{10.62374cm}{\raggedright {\ }{\ }{\ }MeV \textit{25} (\href{https://www.nndc.bnl.gov/nsr/nsrlink.jsp?2024Ko04,B}{2024Ko04}), suggests that the isospin mirror symmetry is\vspace{0.1cm}}&\\
&&&&&&\parbox[t][0.3cm]{10.62374cm}{\raggedright {\ }{\ }{\ }maintained for this state.\vspace{0.1cm}}&\\
\end{longtable}
\begin{textblock}{29}(0,27.3)
Continued on next page (footnotes at end of table)
\end{textblock}
\clearpage
\begin{longtable}{cccccc@{\extracolsep{\fill}}c}
\\[-.4cm]
\multicolumn{7}{c}{{\bf \small \underline{Adopted \hyperlink{8C_LEVEL}{Levels} (continued)}}}\\
\multicolumn{7}{c}{~}\\
\multicolumn{7}{c}{\underline{\ensuremath{^{8}}C Levels (continued)}}\\
\multicolumn{7}{c}{~}\\
\multicolumn{2}{c}{E(level)$^{}$}&\multicolumn{2}{c}{\ensuremath{\Gamma}$^{}$}&XREF&Comments&\\[-.2cm]
\multicolumn{2}{c}{\hrulefill}&\multicolumn{2}{c}{\hrulefill}&\hrulefill&\hrulefill&
\endhead
&&&&&\parbox[t][0.3cm]{11.61438cm}{\raggedright This state is thought to have a 1\textit{p}{\textminus}1\textit{h} configuration based on shell model calculations\vspace{0.1cm}}&\\
&&&&&\parbox[t][0.3cm]{11.61438cm}{\raggedright {\ }{\ }{\ }and a two-body model in (\href{https://www.nndc.bnl.gov/nsr/nsrlink.jsp?2024Ko04,B}{2024Ko04}).\vspace{0.1cm}}&\\
\multicolumn{1}{r@{}}{18.6\ensuremath{\times10^{3}}}&\multicolumn{1}{@{}l}{\ensuremath{^{{\hyperlink{C0LEVEL0}{a}}}} {\it 5}}&\multicolumn{1}{r@{}}{3}&\multicolumn{1}{@{.}l}{9\ensuremath{^{{\hyperlink{C0LEVEL0}{a}}}} MeV {\it 11}}&\multicolumn{1}{l}{\texttt{\hyperlink{C1}{A}\ \ \ \ } }&&\\
\end{longtable}
\parbox[b][0.3cm]{17.7cm}{\makebox[1ex]{\ensuremath{^{\hypertarget{C0LEVEL0}{a}}}} From (\href{https://www.nndc.bnl.gov/nsr/nsrlink.jsp?2011Ch32,B}{2011Ch32}).}\\
\parbox[b][0.3cm]{17.7cm}{\makebox[1ex]{\ensuremath{^{\hypertarget{C0LEVEL1}{b}}}} From the finite-range DWBA analysis with L=1 in (\href{https://www.nndc.bnl.gov/nsr/nsrlink.jsp?2024Ko04,B}{2024Ko04}) using the DWUCK5 computer code. The neutron transfer was}\\
\parbox[b][0.3cm]{17.7cm}{{\ }{\ }considered to be from the 1\textit{p}\ensuremath{_{\textnormal{3/2}}} orbital. Those authors ruled out L=0 for both \ensuremath{^{\textnormal{8}}}C(0, 3.4 MeV) states.}\\
\vspace{0.5cm}
\clearpage
%1H(9C,2H)
\subsection[\hspace{-0.2cm}\ensuremath{^{\textnormal{1}}}H(\ensuremath{^{\textnormal{9}}}C,d)]{ }
\vspace{-27pt}
\vspace{0.3cm}
\hypertarget{C1}{{\bf \small \underline{\ensuremath{^{\textnormal{1}}}H(\ensuremath{^{\textnormal{9}}}C,d)\hspace{0.2in}\href{https://www.nndc.bnl.gov/nsr/nsrlink.jsp?2024Ko04,B}{2024Ko04}}}}\\
\vspace{4pt}
\vspace{8pt}
\parbox[b][0.3cm]{17.7cm}{\addtolength{\parindent}{-0.2in}\href{https://www.nndc.bnl.gov/nsr/nsrlink.jsp?2024Ko04,B}{2024Ko04}: E=55 MeV/nucleon; applied the missing mass technique to reconstruct the excitation energy spectrum of \ensuremath{^{\textnormal{8}}}C (integrated}\\
\parbox[b][0.3cm]{17.7cm}{up to \ensuremath{\theta}\ensuremath{_{\textnormal{c.m.}}}=45\ensuremath{^\circ}) from the \ensuremath{^{\textnormal{1}}}H(\ensuremath{^{\textnormal{9}}}C,d) reaction. A cocktail secondary beam containing 5\% \ensuremath{^{\textnormal{9}}}C at 55 MeV/nucleon, 10\% \ensuremath{^{\textnormal{8}}}B, 65\%}\\
\parbox[b][0.3cm]{17.7cm}{\ensuremath{^{\textnormal{7}}}Be, and 20\% \ensuremath{^{\textnormal{6}}}Li was produced from projectile fragmentation of a 75 MeV/nucleon \ensuremath{^{\textnormal{12}}}C beam on a 2.2-mm Be target at GANIL.}\\
\parbox[b][0.3cm]{17.7cm}{The LISE fragment separator directed this beam to a 1.5-mm-thick cryogenic liquid hydrogen target. Six position sensitive Si-\ensuremath{\Delta}E}\\
\parbox[b][0.3cm]{17.7cm}{backed by CsI(Tl)-E crystals from the MUST2 array, which covered \ensuremath{\theta}\ensuremath{_{\textnormal{lab}}}=2\ensuremath{^\circ}{\textminus}36\ensuremath{^\circ} were used to measure the energies and scattering}\\
\parbox[b][0.3cm]{17.7cm}{angles of the deuterons from the \ensuremath{^{\textnormal{1}}}H(\ensuremath{^{\textnormal{9}}}C,d) reaction. The energy resolution was 40 keV (FWHM) at E\ensuremath{_{\ensuremath{\alpha}}}=5.5 MeV. However, the}\\
\parbox[b][0.3cm]{17.7cm}{uncertain deuterons energy losses through the target limited the experimental energy resolution to 0.50 MeV \textit{5} (r.m.s.).}\\
\parbox[b][0.3cm]{17.7cm}{\addtolength{\parindent}{-0.2in}The excitation function displayed the \ensuremath{^{\textnormal{8}}}C\ensuremath{_{\textnormal{g.s.}}} and two newly observed \ensuremath{^{\textnormal{8}}}C resonances at E\ensuremath{_{\textnormal{x}}}=3.40 MeV \textit{25} (with \ensuremath{\Gamma}=3.0 MeV \textit{5}) and}\\
\parbox[b][0.3cm]{17.7cm}{E\ensuremath{_{\textnormal{x}}}=18.6 MeV \textit{5} (with \ensuremath{\Gamma}=3.9 MeV \textit{11}), which were fitted using Voigt functions convoluted with a background from the simulated}\\
\parbox[b][0.3cm]{17.7cm}{non-resonant breakups due to 1p and 4p emissions. A DWBA analysis was performed using the DWUCK5 computer code to}\\
\parbox[b][0.3cm]{17.7cm}{deduce L and C\ensuremath{^{\textnormal{2}}}S for the \ensuremath{^{\textnormal{8}}}C(0, 3.4 MeV) states. Shell model calculations were performed to obtain J\ensuremath{^{\ensuremath{\pi}}} values from a comparison}\\
\parbox[b][0.3cm]{17.7cm}{of the theoretical and experimental spectroscopic factors. A two-body model was employed to further explore the mirror symmetry}\\
\parbox[b][0.3cm]{17.7cm}{between the unbound 2\ensuremath{^{\textnormal{+}}_{\textnormal{1}}} states in \ensuremath{^{\textnormal{8}}}C and \ensuremath{^{\textnormal{8}}}He.}\\
\vspace{12pt}
\underline{$^{8}$C Levels}\\
\begin{longtable}{ccccccccc@{\extracolsep{\fill}}c}
\multicolumn{2}{c}{E(level)$^{{\hyperlink{C1LEVEL0}{a}}}$}&J$^{\pi}$$^{{\hyperlink{C1LEVEL1}{b}}}$&\multicolumn{2}{c}{\ensuremath{\Gamma}$^{{\hyperlink{C1LEVEL0}{a}}}$}&L$^{{\hyperlink{C1LEVEL1}{b}}}$&\multicolumn{2}{c}{C\ensuremath{^{\textnormal{2}}}S\ensuremath{_{\textnormal{exp}}}$^{{\hyperlink{C1LEVEL1}{b}}}$}&Comments&\\[-.2cm]
\multicolumn{2}{c}{\hrulefill}&\hrulefill&\multicolumn{2}{c}{\hrulefill}&\hrulefill&\multicolumn{2}{c}{\hrulefill}&\hrulefill&
\endfirsthead
\multicolumn{1}{r@{}}{0}&\multicolumn{1}{@{}l}{}&\multicolumn{1}{l}{0\ensuremath{^{+}}}&\multicolumn{1}{r@{}}{130}&\multicolumn{1}{@{ }l}{keV {\it 50}}&\multicolumn{1}{l}{1}&\multicolumn{1}{r@{}}{0}&\multicolumn{1}{@{.}l}{69 {\it 14}}&\parbox[t][0.3cm]{9.680181cm}{\raggedright \ensuremath{\Gamma}: From the \ensuremath{^{\textnormal{8}}}C Adopted Levels and mentioned in (\href{https://www.nndc.bnl.gov/nsr/nsrlink.jsp?2024Ko04,B}{2024Ko04}).\vspace{0.1cm}}&\\
&&&&&&&&\parbox[t][0.3cm]{9.680181cm}{\raggedright C\ensuremath{^{\textnormal{2}}}S\ensuremath{_{\textnormal{theory}}}=\ensuremath{\sim}0.87 from shell model calculations using CK and YSOX\vspace{0.1cm}}&\\
&&&&&&&&\parbox[t][0.3cm]{9.680181cm}{\raggedright {\ }{\ }{\ }interactions for J\ensuremath{^{\ensuremath{\pi}}}=0\ensuremath{^{\textnormal{+}}} (\href{https://www.nndc.bnl.gov/nsr/nsrlink.jsp?2024Ko04,B}{2024Ko04}).\vspace{0.1cm}}&\\
\multicolumn{1}{r@{}}{3.40\ensuremath{\times10^{3}}}&\multicolumn{1}{@{ }l}{{\it 25}}&\multicolumn{1}{l}{2\ensuremath{^{+}}}&\multicolumn{1}{r@{}}{3}&\multicolumn{1}{@{.}l}{0 MeV {\it 5}}&\multicolumn{1}{l}{1}&\multicolumn{1}{r@{}}{0}&\multicolumn{1}{@{.}l}{15 {\it 3}}&\parbox[t][0.3cm]{9.680181cm}{\raggedright T=2 (\href{https://www.nndc.bnl.gov/nsr/nsrlink.jsp?2024Ko04,B}{2024Ko04})\vspace{0.1cm}}&\\
&&&&&&&&\parbox[t][0.3cm]{9.680181cm}{\raggedright \ensuremath{\Gamma}: After the \ensuremath{^{\textnormal{4}}}He*(2\ensuremath{^{\textnormal{+}}_{\textnormal{1}}}) state, this state has the broadest known width\vspace{0.1cm}}&\\
&&&&&&&&\parbox[t][0.3cm]{9.680181cm}{\raggedright {\ }{\ }{\ }among the 2\ensuremath{^{\textnormal{+}}_{\textnormal{1}}} states in even-even nuclei (\href{https://www.nndc.bnl.gov/nsr/nsrlink.jsp?2024Ko04,B}{2024Ko04}).\vspace{0.1cm}}&\\
&&&&&&&&\parbox[t][0.3cm]{9.680181cm}{\raggedright J\ensuremath{^{\pi}}: The L=1 neutron transfer leads to J\ensuremath{^{\ensuremath{\pi}}}=(0, 1, 2)\ensuremath{^{\textnormal{+}}}. However, the\vspace{0.1cm}}&\\
&&&&&&&&\parbox[t][0.3cm]{9.680181cm}{\raggedright {\ }{\ }{\ }resulting experimental spectroscopic factors could only be reproduced\vspace{0.1cm}}&\\
&&&&&&&&\parbox[t][0.3cm]{9.680181cm}{\raggedright {\ }{\ }{\ }using shell model for the case of J\ensuremath{^{\ensuremath{\pi}}}=2\ensuremath{^{\textnormal{+}}} (see below). The other\vspace{0.1cm}}&\\
&&&&&&&&\parbox[t][0.3cm]{9.680181cm}{\raggedright {\ }{\ }{\ }possible assignments would have a negligibly small spectroscopic\vspace{0.1cm}}&\\
&&&&&&&&\parbox[t][0.3cm]{9.680181cm}{\raggedright {\ }{\ }{\ }factors (\href{https://www.nndc.bnl.gov/nsr/nsrlink.jsp?2024Ko04,B}{2024Ko04}).\vspace{0.1cm}}&\\
&&&&&&&&\parbox[t][0.3cm]{9.680181cm}{\raggedright C\ensuremath{^{\textnormal{2}}}S\ensuremath{_{\textnormal{theory}}}=0.13 from shell model calculations using CK and YSOX\vspace{0.1cm}}&\\
&&&&&&&&\parbox[t][0.3cm]{9.680181cm}{\raggedright {\ }{\ }{\ }interactions for J\ensuremath{^{\ensuremath{\pi}}}=2\ensuremath{^{\textnormal{+}}} (\href{https://www.nndc.bnl.gov/nsr/nsrlink.jsp?2024Ko04,B}{2024Ko04}). So, those authors recommended\vspace{0.1cm}}&\\
&&&&&&&&\parbox[t][0.3cm]{9.680181cm}{\raggedright {\ }{\ }{\ }J\ensuremath{^{\ensuremath{\pi}}}=2\ensuremath{^{\textnormal{+}}}.\vspace{0.1cm}}&\\
&&&&&&&&\parbox[t][0.3cm]{9.680181cm}{\raggedright This state is thought to have a 1\textit{p}{\textminus}1\textit{h} configuration based on shell\vspace{0.1cm}}&\\
&&&&&&&&\parbox[t][0.3cm]{9.680181cm}{\raggedright {\ }{\ }{\ }model calculations and a two-body model (\href{https://www.nndc.bnl.gov/nsr/nsrlink.jsp?2024Ko04,B}{2024Ko04}).\vspace{0.1cm}}&\\
\multicolumn{1}{r@{}}{18.6\ensuremath{\times10^{3}}}&\multicolumn{1}{@{ }l}{{\it 5}}&&\multicolumn{1}{r@{}}{3}&\multicolumn{1}{@{.}l}{9 MeV {\it 11}}&&&&&\\
\end{longtable}
\parbox[b][0.3cm]{17.7cm}{\makebox[1ex]{\ensuremath{^{\hypertarget{C1LEVEL0}{a}}}} From (\href{https://www.nndc.bnl.gov/nsr/nsrlink.jsp?2024Ko04,B}{2024Ko04}) unless otherwise noted.}\\
\parbox[b][0.3cm]{17.7cm}{\makebox[1ex]{\ensuremath{^{\hypertarget{C1LEVEL1}{b}}}} From the finite-range DWBA analysis in (\href{https://www.nndc.bnl.gov/nsr/nsrlink.jsp?2024Ko04,B}{2024Ko04}) using the DWUCK5 computer code. The neutron transfer was considered to}\\
\parbox[b][0.3cm]{17.7cm}{{\ }{\ }be from the 1\textit{p}\ensuremath{_{\textnormal{3/2}}} orbital. Those authors ruled out L=0 for both \ensuremath{^{\textnormal{8}}}C(0, 3.4 MeV) states.}\\
\vspace{0.5cm}
\clearpage
%9BE(9C,8C)
\subsection[\hspace{-0.2cm}\ensuremath{^{\textnormal{9}}}Be(\ensuremath{^{\textnormal{9}}}C,\ensuremath{^{\textnormal{8}}}C)]{ }
\vspace{-27pt}
\vspace{0.3cm}
\hypertarget{C2}{{\bf \small \underline{\ensuremath{^{\textnormal{9}}}Be(\ensuremath{^{\textnormal{9}}}C,\ensuremath{^{\textnormal{8}}}C)\hspace{0.2in}\href{https://www.nndc.bnl.gov/nsr/nsrlink.jsp?2010Ch42,B}{2010Ch42},\href{https://www.nndc.bnl.gov/nsr/nsrlink.jsp?2011Ch32,B}{2011Ch32}}}}\\
\vspace{4pt}
\vspace{8pt}
\parbox[b][0.3cm]{17.7cm}{\addtolength{\parindent}{-0.2in}\href{https://www.nndc.bnl.gov/nsr/nsrlink.jsp?2010Ch42,B}{2010Ch42}: The authors measured the multi-proton decay properties of \ensuremath{^{\textnormal{8}}}C by measuring the complete kinematics of remnant \ensuremath{\alpha}+4p}\\
\parbox[b][0.3cm]{17.7cm}{decay products. The proton correlations indicate that the decay follows a \ensuremath{^{\textnormal{8}}}C\ensuremath{\rightarrow}\ensuremath{^{\textnormal{6}}}Be+2p\ensuremath{\rightarrow}\ensuremath{\alpha}+2p+2p multi{\textminus}step path. A beam of 70}\\
\parbox[b][0.3cm]{17.7cm}{MeV/nucleon \ensuremath{^{\textnormal{9}}}C was produced by fragmentation of an \ensuremath{^{\textnormal{16}}}O beam at the NSCL. The \ensuremath{^{\textnormal{9}}}C beam impinged on a \ensuremath{^{\textnormal{9}}}Be target and short}\\
\parbox[b][0.3cm]{17.7cm}{lived unbound nuclei produced in the reactions were studied by reconstruction of the breakup particle kinematics. The proton-proton}\\
\parbox[b][0.3cm]{17.7cm}{pairing correlations indicate that 92\% \textit{5} of events proceed through the 2p+\ensuremath{^{\textnormal{6}}}Be\ensuremath{_{\textnormal{g.s.}}} decay channel. Combined with results on \ensuremath{^{\textnormal{6}}}Be,}\\
\parbox[b][0.3cm]{17.7cm}{this indicates a \ensuremath{^{\textnormal{8}}}C\ensuremath{\rightarrow}\ensuremath{^{\textnormal{6}}}Be+2p\ensuremath{\rightarrow}\ensuremath{\alpha}+2p+2p decay path. Data on \ensuremath{^{\textnormal{8}}}B* decay is consistent with 2p decay from \ensuremath{^{\textnormal{8}}}B*(10.61 MeV) which is}\\
\parbox[b][0.3cm]{17.7cm}{the IAS of \ensuremath{^{\textnormal{8}}}C\ensuremath{_{\textnormal{g.s.}}}.}\\
\parbox[b][0.3cm]{17.7cm}{\addtolength{\parindent}{-0.2in}\href{https://www.nndc.bnl.gov/nsr/nsrlink.jsp?2011Ch32,B}{2011Ch32}: The authors impinged a 70 MeV/nucleon \ensuremath{^{\textnormal{9}}}C beam on a thick \ensuremath{^{\textnormal{9}}}Be target and detected ejected reaction products with a}\\
\parbox[b][0.3cm]{17.7cm}{large area position sensitive \ensuremath{\Delta}E-E array. Reconstruction of the complete kinematics permitted an analysis of excitation energies,}\\
\parbox[b][0.3cm]{17.7cm}{decay pathways and associated branching ratios for several nuclei. A beam of 150 MeV/nucleon \ensuremath{^{\textnormal{16}}}O ions was fragmented on a}\\
\parbox[b][0.3cm]{17.7cm}{thick \ensuremath{^{\textnormal{9}}}Be target to produce a 70 MeV/nucleon \ensuremath{^{\textnormal{9}}}C beam using the A1900 fragment separator at the NSCL. The \ensuremath{^{\textnormal{9}}}C beam impinged}\\
\parbox[b][0.3cm]{17.7cm}{on a 1mm thick \ensuremath{^{\textnormal{9}}}Be target and reaction products were detected in 14 position sensitive \ensuremath{\Delta}E-E detectors of the HiRA array. The}\\
\parbox[b][0.3cm]{17.7cm}{coincident reaction products were analyzed via kinematic energy reconstruction to evaluate excitation energies and decay paths. The}\\
\parbox[b][0.3cm]{17.7cm}{authors obtained the \ensuremath{^{\textnormal{8}}}C mass excess \ensuremath{\Delta}M(\ensuremath{^{\textnormal{8}}}C)=35.030 MeV \textit{30} and the width \ensuremath{\Gamma}=130 keV \textit{50}.}\\
\vspace{12pt}
\underline{$^{8}$C Levels}\\
\begin{longtable}{ccccc@{\extracolsep{\fill}}c}
\multicolumn{2}{c}{E(level)$^{}$}&\multicolumn{2}{c}{\ensuremath{\Gamma}$^{{\hyperlink{C2LEVEL0}{a}}}$}&Comments&\\[-.2cm]
\multicolumn{2}{c}{\hrulefill}&\multicolumn{2}{c}{\hrulefill}&\hrulefill&
\endfirsthead
\multicolumn{1}{r@{}}{0}&\multicolumn{1}{@{}l}{}&\multicolumn{1}{r@{}}{130}&\multicolumn{1}{@{ }l}{keV {\it 50}}&\parbox[t][0.3cm]{13.76276cm}{\raggedright (\href{https://www.nndc.bnl.gov/nsr/nsrlink.jsp?2011Ch32,B}{2011Ch32}) obtained mass excess 35.030 MeV \textit{30}.\vspace{0.1cm}}&\\
\end{longtable}
\parbox[b][0.3cm]{17.7cm}{\makebox[1ex]{\ensuremath{^{\hypertarget{C2LEVEL0}{a}}}} From (\href{https://www.nndc.bnl.gov/nsr/nsrlink.jsp?2011Ch32,B}{2011Ch32}).}\\
\vspace{0.5cm}
\clearpage
%9BE(13O,8C)
\subsection[\hspace{-0.2cm}\ensuremath{^{\textnormal{9}}}Be(\ensuremath{^{\textnormal{13}}}O,\ensuremath{^{\textnormal{8}}}C)]{ }
\vspace{-27pt}
\vspace{0.3cm}
\hypertarget{C3}{{\bf \small \underline{\ensuremath{^{\textnormal{9}}}Be(\ensuremath{^{\textnormal{13}}}O,\ensuremath{^{\textnormal{8}}}C)\hspace{0.2in}\href{https://www.nndc.bnl.gov/nsr/nsrlink.jsp?2023Ch46,B}{2023Ch46}}}}\\
\vspace{4pt}
\vspace{8pt}
\parbox[b][0.3cm]{17.7cm}{\addtolength{\parindent}{-0.2in}\href{https://www.nndc.bnl.gov/nsr/nsrlink.jsp?2023Ch46,B}{2023Ch46}: XUNDL file compiled by TUNL (2023).}\\
\parbox[b][0.3cm]{17.7cm}{\addtolength{\parindent}{-0.2in}The authors report first evidence for states in \ensuremath{^{\textnormal{9}}}N; the states are found to decay via 1-proton decay to \ensuremath{^{\textnormal{8}}}C\ensuremath{_{\textnormal{g.s.}}}.}\\
\parbox[b][0.3cm]{17.7cm}{\addtolength{\parindent}{-0.2in}The measurement utilized a beam of \ensuremath{\approx}60 MeV/nucleon \ensuremath{^{\textnormal{13}}}O ions from the NSCL/A1900 fragment separator that was purified in the}\\
\parbox[b][0.3cm]{17.7cm}{Radio Frequency Fragment Separator before impinging on a 1 mm thick \ensuremath{^{\textnormal{9}}}Be target (see supplemental material and prior reports}\\
\parbox[b][0.3cm]{17.7cm}{such as (\href{https://www.nndc.bnl.gov/nsr/nsrlink.jsp?2023Ch22,B}{2023Ch22})). Fragmentation reactions populated the short-lived \ensuremath{^{\textnormal{9}}}N nucleus, which proton decayed before exiting the target.}\\
\parbox[b][0.3cm]{17.7cm}{The complete kinematics of the charged-particle reaction products were measured using the HiRA array, which comprised a set of}\\
\parbox[b][0.3cm]{17.7cm}{14 64 mm \ensuremath{\times} 64 mm position sensitive \ensuremath{\Delta}E-E telescopes that covered the forward direction of the outgoing beam (\ensuremath{\theta}\ensuremath{_{\textnormal{lab}}}\ensuremath{\approx}2.1\ensuremath{^\circ} to}\\
\parbox[b][0.3cm]{17.7cm}{12.4\ensuremath{^\circ}). We assume the telescopes were arranged in vertical towers with a 2-3-4-3-2 configuration where the central tower had a}\\
\parbox[b][0.3cm]{17.7cm}{gap between the upper and lower two telescopes to permit the beam a downstream exit at \ensuremath{\theta}=0\ensuremath{^\circ}, as in past experiments.}\\
\parbox[b][0.3cm]{17.7cm}{\addtolength{\parindent}{-0.2in}Part of the analysis of the \ensuremath{^{\textnormal{9}}}N properties required an independent analysis of the \ensuremath{^{\textnormal{8}}}C ground state properties where the 4p+\ensuremath{\alpha}}\\
\parbox[b][0.3cm]{17.7cm}{sub-events were analyzed by fitting the data with a Breit-Wigner lineshape and a smooth background component. A \ensuremath{^{\textnormal{8}}}C\ensuremath{_{\textnormal{g.s.}}} peak}\\
\parbox[b][0.3cm]{17.7cm}{was found with E\ensuremath{_{\textnormal{c.m.}}}(4p+\ensuremath{\alpha})=3490 keV \textit{20} and \ensuremath{\Gamma}=88 keV \textit{61} (supplemental materials). \textit{Note:}\ensuremath{^{\textnormal{8}}}C is known to decay sequentially by}\\
\parbox[b][0.3cm]{17.7cm}{two 2-proton decays having an intermediate state involving \ensuremath{^{\textnormal{6}}}Be\ensuremath{_{\textnormal{g.s.}}}; \ensuremath{^{\textnormal{8}}}C\ensuremath{\rightarrow}2p+[\ensuremath{^{\textnormal{6}}}Be\ensuremath{_{\textnormal{g.s.}}}\ensuremath{\rightarrow}2p+\ensuremath{^{\textnormal{4}}}He].}\\
\vspace{12pt}
\underline{$^{8}$C Levels}\\
\begin{longtable}{cccccc@{\extracolsep{\fill}}c}
\multicolumn{2}{c}{E(level)$^{}$}&J$^{\pi}$$^{}$&\multicolumn{2}{c}{T$_{1/2}$$^{}$}&Comments&\\[-.2cm]
\multicolumn{2}{c}{\hrulefill}&\hrulefill&\multicolumn{2}{c}{\hrulefill}&\hrulefill&
\endfirsthead
\multicolumn{1}{r@{}}{0}&\multicolumn{1}{@{}l}{}&\multicolumn{1}{l}{0\ensuremath{^{+}}}&\multicolumn{1}{r@{}}{88}&\multicolumn{1}{@{ }l}{keV {\it 61}}&\parbox[t][0.3cm]{13.1808605cm}{\raggedright E(level): From E\ensuremath{_{\textnormal{c.m.}}}(4p+\ensuremath{\alpha})=3490 keV \textit{20} (\href{https://www.nndc.bnl.gov/nsr/nsrlink.jsp?2023Ch46,B}{2023Ch46}). From this energy, the mass of \ensuremath{^{\textnormal{8}}}C was\vspace{0.1cm}}&\\
&&&&&\parbox[t][0.3cm]{13.1808605cm}{\raggedright {\ }{\ }{\ }deduced to be 8.03765 u \textit{16}.\vspace{0.1cm}}&\\
&&&&&\parbox[t][0.3cm]{13.1808605cm}{\raggedright T\ensuremath{_{1/2}}: From (\href{https://www.nndc.bnl.gov/nsr/nsrlink.jsp?2011Ch32,B}{2011Ch32}).\vspace{0.1cm}}&\\
&&&&&\parbox[t][0.3cm]{13.1808605cm}{\raggedright The decay proceeds by \ensuremath{^{\textnormal{8}}}C\ensuremath{\rightarrow}2p+\ensuremath{^{\textnormal{6}}}Be\ensuremath{_{\textnormal{g.s.}}}\ensuremath{\rightarrow}4p+\ensuremath{^{\textnormal{4}}}He (\href{https://www.nndc.bnl.gov/nsr/nsrlink.jsp?2023Ch46,B}{2023Ch46}).\vspace{0.1cm}}&\\
\end{longtable}
\clearpage
%12C(A,8HE)
\subsection[\hspace{-0.2cm}\ensuremath{^{\textnormal{12}}}C(\ensuremath{\alpha},\ensuremath{^{\textnormal{8}}}He)]{ }
\vspace{-27pt}
\vspace{0.3cm}
\hypertarget{C4}{{\bf \small \underline{\ensuremath{^{\textnormal{12}}}C(\ensuremath{\alpha},\ensuremath{^{\textnormal{8}}}He)\hspace{0.2in}\href{https://www.nndc.bnl.gov/nsr/nsrlink.jsp?1974Ro17,B}{1974Ro17},\href{https://www.nndc.bnl.gov/nsr/nsrlink.jsp?1976Tr01,B}{1976Tr01}}}}\\
\vspace{4pt}
\vspace{8pt}
\parbox[b][0.3cm]{17.7cm}{\addtolength{\parindent}{-0.2in}\href{https://www.nndc.bnl.gov/nsr/nsrlink.jsp?1974Ro17,B}{1974Ro17}: E=156 MeV; measured Q of \ensuremath{^{\textnormal{8}}}He spectrum, \ensuremath{\sigma}, deduced \ensuremath{^{\textnormal{8}}}C mass excess and width. This is the article in which \ensuremath{^{\textnormal{8}}}C is}\\
\parbox[b][0.3cm]{17.7cm}{first recognized (\href{https://www.nndc.bnl.gov/nsr/nsrlink.jsp?2012Th01,B}{2012Th01}). The differential cross section was found to be about 20 nb/sr at \ensuremath{\theta}\ensuremath{_{\textnormal{lab}}}=2\ensuremath{^\circ}. The mass excess of \ensuremath{^{\textnormal{8}}}C was}\\
\parbox[b][0.3cm]{17.7cm}{found to be \ensuremath{\Delta}M(\ensuremath{^{\textnormal{8}}}C)=35.30 MeV \textit{20}. As indicated above, \ensuremath{^{\textnormal{8}}}C decays by proton emission. Assuming a Gaussian line shape, the}\\
\parbox[b][0.3cm]{17.7cm}{width of the observed \ensuremath{^{\textnormal{8}}}C state is found to be \ensuremath{\Gamma}=0.22 MeV \textit{+8{\textminus}14}.}\\
\parbox[b][0.3cm]{17.7cm}{\addtolength{\parindent}{-0.2in}Since the \ensuremath{^{\textnormal{8}}}He spectrum is the observed quantity in this experiment, a change in the measured mass of \ensuremath{^{\textnormal{8}}}He would lead to a change}\\
\parbox[b][0.3cm]{17.7cm}{in the mass of \ensuremath{^{\textnormal{8}}}C. In (\href{https://www.nndc.bnl.gov/nsr/nsrlink.jsp?1974Ce05,B}{1974Ce05}) a more accurate value of the mass defect of \ensuremath{^{\textnormal{8}}}He led to a revision of the measured mass defect of}\\
\parbox[b][0.3cm]{17.7cm}{\ensuremath{^{\textnormal{8}}}C, \ensuremath{\Delta}M(\ensuremath{^{\textnormal{8}}}C)=35.38 MeV \textit{17}.}\\
\parbox[b][0.3cm]{17.7cm}{\addtolength{\parindent}{-0.2in}\href{https://www.nndc.bnl.gov/nsr/nsrlink.jsp?1976Tr01,B}{1976Tr01}: E=123.5 MeV; measured \ensuremath{\sigma}, deduced mass excess and width. The mass excess of \ensuremath{^{\textnormal{8}}}C was found to be \ensuremath{\Delta}M(\ensuremath{^{\textnormal{8}}}C)=35.10}\\
\parbox[b][0.3cm]{17.7cm}{MeV \textit{3}. The width was found to be \ensuremath{\Gamma}=230 keV \textit{50} assuming a Gaussian fit and 183 keV \textit{56} assuming a Breit-Wigner fit. An}\\
\parbox[b][0.3cm]{17.7cm}{IMME study of A=8 nuclei is reported in this article.}\\
\vspace{12pt}
\underline{$^{8}$C Levels}\\
\begin{longtable}{ccccc@{\extracolsep{\fill}}c}
\multicolumn{2}{c}{E(level)$^{}$}&\multicolumn{2}{c}{\ensuremath{\Gamma}$^{}$}&Comments&\\[-.2cm]
\multicolumn{2}{c}{\hrulefill}&\multicolumn{2}{c}{\hrulefill}&\hrulefill&
\endfirsthead
\multicolumn{1}{r@{}}{0}&\multicolumn{1}{@{}l}{}&\multicolumn{1}{r@{}}{230}&\multicolumn{1}{@{ }l}{keV {\it 50}}&\parbox[t][0.3cm]{13.76276cm}{\raggedright \ensuremath{\Gamma}: From (\href{https://www.nndc.bnl.gov/nsr/nsrlink.jsp?1976Tr01,B}{1976Tr01}). See also \ensuremath{\Gamma}=0.22 MeV \textit{+8{\textminus}14} (\href{https://www.nndc.bnl.gov/nsr/nsrlink.jsp?1974Ro17,B}{1974Ro17}).\vspace{0.1cm}}&\\
&&&&\parbox[t][0.3cm]{13.76276cm}{\raggedright (\href{https://www.nndc.bnl.gov/nsr/nsrlink.jsp?1976Tr01,B}{1976Tr01}) obtained mass excess 35.10 MeV \textit{3}.\vspace{0.1cm}}&\\
\end{longtable}
\clearpage
%14N(3HE,9LI)
\subsection[\hspace{-0.2cm}\ensuremath{^{\textnormal{14}}}N(\ensuremath{^{\textnormal{3}}}He,\ensuremath{^{\textnormal{9}}}Li)]{ }
\vspace{-27pt}
\vspace{0.3cm}
\hypertarget{C5}{{\bf \small \underline{\ensuremath{^{\textnormal{14}}}N(\ensuremath{^{\textnormal{3}}}He,\ensuremath{^{\textnormal{9}}}Li)\hspace{0.2in}\href{https://www.nndc.bnl.gov/nsr/nsrlink.jsp?1976Ro04,B}{1976Ro04}}}}\\
\vspace{4pt}
\vspace{8pt}
\parbox[b][0.3cm]{17.7cm}{\addtolength{\parindent}{-0.2in}\href{https://www.nndc.bnl.gov/nsr/nsrlink.jsp?1976Ro04,B}{1976Ro04}: The \ensuremath{^{\textnormal{3}}}He beam with energy E=76 MeV from the MSU cyclotron collided with a target of either a solid melamine}\\
\parbox[b][0.3cm]{17.7cm}{(C\ensuremath{_{\textnormal{3}}}N\ensuremath{_{\textnormal{6}}}H\ensuremath{_{\textnormal{6}}}) or N\ensuremath{_{\textnormal{2}}} gas and the \ensuremath{^{\textnormal{9}}}Li spectrum was observed. Measured laboratory cross sections with approximately 40\% uncertainties}\\
\parbox[b][0.3cm]{17.7cm}{are d\ensuremath{\sigma}/d\ensuremath{\Omega}=3 nb/sr and 5 nb/sr at \ensuremath{\theta}\ensuremath{_{\textnormal{lab}}}=8\ensuremath{^\circ} and 10\ensuremath{^\circ}, respectively. The authors determined the mass excess of \ensuremath{^{\textnormal{8}}}C to be}\\
\parbox[b][0.3cm]{17.7cm}{\ensuremath{\Delta}M(\ensuremath{^{\textnormal{8}}}C)=35.06 MeV \textit{5}. Assuming a Gaussian shape for the line shape, the width was found to be \ensuremath{\Gamma}=290 keV \textit{80}.}\\
\vspace{12pt}
\underline{$^{8}$C Levels}\\
\begin{longtable}{ccccc@{\extracolsep{\fill}}c}
\multicolumn{2}{c}{E(level)$^{}$}&\multicolumn{2}{c}{\ensuremath{\Gamma}$^{}$}&Comments&\\[-.2cm]
\multicolumn{2}{c}{\hrulefill}&\multicolumn{2}{c}{\hrulefill}&\hrulefill&
\endfirsthead
\multicolumn{1}{r@{}}{0}&\multicolumn{1}{@{}l}{}&\multicolumn{1}{r@{}}{290}&\multicolumn{1}{@{ }l}{keV {\it 80}}&\parbox[t][0.3cm]{13.76276cm}{\raggedright (\href{https://www.nndc.bnl.gov/nsr/nsrlink.jsp?1976Ro04,B}{1976Ro04}) obtained mass excess 35.06 MeV \textit{5}.\vspace{0.1cm}}&\\
\end{longtable}
\end{center}
\clearpage
\newpage
\pagestyle{plain}
\section[References]{ }
\vspace{-30pt}
\begin{longtable}{l@{\hskip 0.9cm}l}
\multicolumn{2}{c}{REFERENCES FOR A=8}\\
&\endfirsthead
\multicolumn{2}{c}{REFERENCES FOR A=8(CONTINUED)}\\
&\endhead
\href{https://www.nndc.bnl.gov/nsr/nsrlink.jsp?1974Ce05,B}{1974Ce05}&\parbox[t]{6in}{\addtolength{\parindent}{-0.25cm}J.Cerny, N.A.Jelley, D.L.Hendrie, C.F.Maguire et al. - Phys.Rev. C10, 2654 (1974).}\\
&\parbox[t]{6in}{\addtolength{\parindent}{-0.25cm} \textit{A More Accurate Mass for \ensuremath{^{\textnormal{8}}}He.}}\\
\href{https://www.nndc.bnl.gov/nsr/nsrlink.jsp?1974Ir04,B}{1974Ir04}&\parbox[t]{6in}{\addtolength{\parindent}{-0.25cm}J.M.Irvine, G.S.Mani, M.Vallieres - Czech.J.Phys. 24B, 1269 (1974).}\\
&\parbox[t]{6in}{\addtolength{\parindent}{-0.25cm} \textit{The Structure of Light p-Shell Nuclei.}}\\
\href{https://www.nndc.bnl.gov/nsr/nsrlink.jsp?1974Ro17,B}{1974Ro17}&\parbox[t]{6in}{\addtolength{\parindent}{-0.25cm}R.G.H.Robertson, S.Martin, W.R.Falk, D.Ingham, A.Djaloeis - Phys.Rev.Lett. 32, 1207 (1974).}\\
&\parbox[t]{6in}{\addtolength{\parindent}{-0.25cm} \textit{Highly Proton-Rich T(z) = {\textminus}2 Nuclides: \ensuremath{^{\textnormal{8}}}C and \ensuremath{^{\textnormal{20}}}Mg.}}\\
\href{https://www.nndc.bnl.gov/nsr/nsrlink.jsp?1976Ro04,B}{1976Ro04}&\parbox[t]{6in}{\addtolength{\parindent}{-0.25cm}R.G.H.Robertson, W.Benenson, E.Kashy, D.Mueller - Phys.Rev. C13, 1018 (1976).}\\
&\parbox[t]{6in}{\addtolength{\parindent}{-0.25cm} \textit{Measurement of the Mass of \ensuremath{^{\textnormal{8}}}C by the \ensuremath{^{\textnormal{14}}}N(\ensuremath{^{\textnormal{3}}}He,\ensuremath{^{\textnormal{9}}}Li) Reaction.}}\\
\href{https://www.nndc.bnl.gov/nsr/nsrlink.jsp?1976Tr01,B}{1976Tr01}&\parbox[t]{6in}{\addtolength{\parindent}{-0.25cm}R.E.Tribble, R.A.Kenefick, R.L.Spross - Phys.Rev. C13, 50 (1976).}\\
&\parbox[t]{6in}{\addtolength{\parindent}{-0.25cm} \textit{Mass of \ensuremath{^{\textnormal{8}}}C.}}\\
\href{https://www.nndc.bnl.gov/nsr/nsrlink.jsp?1984An18,B}{1984An18}&\parbox[t]{6in}{\addtolength{\parindent}{-0.25cm}M.S.Antony, A.Pape - Phys.Rev. C30, 1286 (1984).}\\
&\parbox[t]{6in}{\addtolength{\parindent}{-0.25cm} \textit{Isobaric Mass Systematics for A \ensuremath{\leq} 60.}}\\
\href{https://www.nndc.bnl.gov/nsr/nsrlink.jsp?1987Bl18,B}{1987Bl18}&\parbox[t]{6in}{\addtolength{\parindent}{-0.25cm}R.Blumel, K.Dietrich - Nucl.Phys. A471, 453 (1987).}\\
&\parbox[t]{6in}{\addtolength{\parindent}{-0.25cm} \textit{Excited States of Light N = Z Nuclei with a Specific Spin-Isospin Order.}}\\
\href{https://www.nndc.bnl.gov/nsr/nsrlink.jsp?1987Sa15,B}{1987Sa15}&\parbox[t]{6in}{\addtolength{\parindent}{-0.25cm}H.Sagawa, H.Toki - J.Phys.(London) G13, 453 (1987).}\\
&\parbox[t]{6in}{\addtolength{\parindent}{-0.25cm} \textit{Hartree-Fock Calculations of Light Neutron-Rich Nuclei.}}\\
\href{https://www.nndc.bnl.gov/nsr/nsrlink.jsp?1988Co15,B}{1988Co15}&\parbox[t]{6in}{\addtolength{\parindent}{-0.25cm}E.Comay, I.Kelson, A.Zidon - Phys.Lett. 210B, 31 (1988).}\\
&\parbox[t]{6in}{\addtolength{\parindent}{-0.25cm} \textit{The Thomas-Ehrman Shift across the Proton Dripline.}}\\
\href{https://www.nndc.bnl.gov/nsr/nsrlink.jsp?1996Gr21,B}{1996Gr21}&\parbox[t]{6in}{\addtolength{\parindent}{-0.25cm}F.Grummer, B.Q.Chen, Z.Y.Ma, S.Krewald - Phys.Lett. 387B, 673 (1996).}\\
&\parbox[t]{6in}{\addtolength{\parindent}{-0.25cm} \textit{Bulk Properties of Light Deformed Nuclei Derived from a Medium-Modified Meson-Exchange Interaction.}}\\
\href{https://www.nndc.bnl.gov/nsr/nsrlink.jsp?1996Ka14,B}{1996Ka14}&\parbox[t]{6in}{\addtolength{\parindent}{-0.25cm}Y.Kanada-Enyo, H.Horiuchi - Phys.Rev. C54, R468 (1996).}\\
&\parbox[t]{6in}{\addtolength{\parindent}{-0.25cm} \textit{Magnetic Moments of C Isotopes Studied with Antisymmetrized MolecularDynamics.}}\\
\href{https://www.nndc.bnl.gov/nsr/nsrlink.jsp?1996Su24,B}{1996Su24}&\parbox[t]{6in}{\addtolength{\parindent}{-0.25cm}Y.Sugahara, K.Sumiyoshi, H.Toki, A.Ozawa, I.Tanihata - Prog.Theor.Phys.(Kyoto) 96, 1165 (1996).}\\
&\parbox[t]{6in}{\addtolength{\parindent}{-0.25cm} \textit{Study of Light Nuclei in the Relativistic Mean Field Theory and the Non-Relativistic Skyrme-Hartree-Fock Theory.}}\\
\href{https://www.nndc.bnl.gov/nsr/nsrlink.jsp?1997Ba54,B}{1997Ba54}&\parbox[t]{6in}{\addtolength{\parindent}{-0.25cm}X.Bai, J.Hu - Phys.Rev. C56, 1410 (1997).}\\
&\parbox[t]{6in}{\addtolength{\parindent}{-0.25cm} \textit{Microscopic Study of the Ground State Properties of Light Nuclei.}}\\
\href{https://www.nndc.bnl.gov/nsr/nsrlink.jsp?1997Po12,B}{1997Po12}&\parbox[t]{6in}{\addtolength{\parindent}{-0.25cm}I.V.Poplavsky, M.N.Popushoi - Bull.Rus.Acad.Sci.Phys. 61, 160 (1997).}\\
&\parbox[t]{6in}{\addtolength{\parindent}{-0.25cm} \textit{Coulomb Energies of Light Nuclei.}}\\
\href{https://www.nndc.bnl.gov/nsr/nsrlink.jsp?1998Br09,B}{1998Br09}&\parbox[t]{6in}{\addtolength{\parindent}{-0.25cm}J.Britz, A.Pape, M.S.Anthony - At.Data Nucl.Data Tables 69, 125 (1998).}\\
&\parbox[t]{6in}{\addtolength{\parindent}{-0.25cm} \textit{Coefficients of the Isobaric Mass Equation and Their Correlations withVarious Nuclear Parameters.}}\\
\href{https://www.nndc.bnl.gov/nsr/nsrlink.jsp?1998Wi10,B}{1998Wi10}&\parbox[t]{6in}{\addtolength{\parindent}{-0.25cm}R.B.Wiringa - Nucl.Phys. A631, 70c (1998).}\\
&\parbox[t]{6in}{\addtolength{\parindent}{-0.25cm} \textit{Quantum Monte Carlo Calculations for Light Nuclei.}}\\
\href{https://www.nndc.bnl.gov/nsr/nsrlink.jsp?1999Ha61,B}{1999Ha61}&\parbox[t]{6in}{\addtolength{\parindent}{-0.25cm}I.Hamamoto, H.Sagawa - Phys.Rev. C60, 064314 (1999).}\\
&\parbox[t]{6in}{\addtolength{\parindent}{-0.25cm} \textit{Response of Light Drip Line Nuclei to Spin Dependent Operators.}}\\
\href{https://www.nndc.bnl.gov/nsr/nsrlink.jsp?2000Wi09,B}{2000Wi09}&\parbox[t]{6in}{\addtolength{\parindent}{-0.25cm}R.B.Wiringa, S.C.Pieper, J.Carlson, V.R.Pandharipande - Phys.Rev. C62, 014001 (2000).}\\
&\parbox[t]{6in}{\addtolength{\parindent}{-0.25cm} \textit{Quantum Monte Carlo Calculations of A = 8 Nuclei.}}\\
\href{https://www.nndc.bnl.gov/nsr/nsrlink.jsp?2001Co21,B}{2001Co21}&\parbox[t]{6in}{\addtolength{\parindent}{-0.25cm}L.Coraggio, A.Covello, A.Gargano, N.Itaco, T.T.S.Kuo - J.Phys.(London) G27, 2351 (2001).}\\
&\parbox[t]{6in}{\addtolength{\parindent}{-0.25cm} \textit{Two-Frequency Shell-Model Calculations for p-Shell Nuclei.}}\\
\href{https://www.nndc.bnl.gov/nsr/nsrlink.jsp?2002Ba90,B}{2002Ba90}&\parbox[t]{6in}{\addtolength{\parindent}{-0.25cm}F.C.Barker - Phys.Rev. C66, 047603 (2002); Erratum Phys.Rev. C67, 049902 (2003).}\\
&\parbox[t]{6in}{\addtolength{\parindent}{-0.25cm} \textit{\ensuremath{^{\textnormal{6}}}Be and \ensuremath{^{\textnormal{8}}}C level widths.}}\\
\href{https://www.nndc.bnl.gov/nsr/nsrlink.jsp?2002Fo11,B}{2002Fo11}&\parbox[t]{6in}{\addtolength{\parindent}{-0.25cm}H.T.Fortune, R.Sherr - Phys.Rev. C66, 017301 (2002).}\\
&\parbox[t]{6in}{\addtolength{\parindent}{-0.25cm} \textit{Structure of \ensuremath{^{\textnormal{16}}}Ne Ground State.}}\\
\href{https://www.nndc.bnl.gov/nsr/nsrlink.jsp?2003Ba99,B}{2003Ba99}&\parbox[t]{6in}{\addtolength{\parindent}{-0.25cm}F.C.Barker - Phys.Rev. C 68, 054602 (2003).}\\
&\parbox[t]{6in}{\addtolength{\parindent}{-0.25cm} \textit{R-matrix formulas for three-body decay widths.}}\\
\href{https://www.nndc.bnl.gov/nsr/nsrlink.jsp?2006Sa29,B}{2006Sa29}&\parbox[t]{6in}{\addtolength{\parindent}{-0.25cm}P.Saviankou, F.Grummer, E.Epelbaum, S.Krewald, U.-G.Meissner - Phys.Atomic Nuclei 69, 1119 (2006).}\\
&\parbox[t]{6in}{\addtolength{\parindent}{-0.25cm} \textit{Effective Field Theory Approach to Nuclear Matter.}}\\
\href{https://www.nndc.bnl.gov/nsr/nsrlink.jsp?2006Wi07,B}{2006Wi07}&\parbox[t]{6in}{\addtolength{\parindent}{-0.25cm}R.B.Wiringa - Phys.Rev. C 73, 034317 (2006).}\\
&\parbox[t]{6in}{\addtolength{\parindent}{-0.25cm} \textit{Pair counting, pion-exchange forces and the structure of light nuclei.}}\\
\href{https://www.nndc.bnl.gov/nsr/nsrlink.jsp?2007Ma79,B}{2007Ma79}&\parbox[t]{6in}{\addtolength{\parindent}{-0.25cm}O.V.Manko, N.S.Manton, S.W.Wood - Phys.Rev. C 76, 055203 (2007).}\\
&\parbox[t]{6in}{\addtolength{\parindent}{-0.25cm} \textit{Light nuclei as quantized Skyrmions.}}\\
\href{https://www.nndc.bnl.gov/nsr/nsrlink.jsp?2009Ba41,B}{2009Ba41}&\parbox[t]{6in}{\addtolength{\parindent}{-0.25cm}R.A.Battye, N.S.Manton, P.M.Sutcliffe, S.W.Wood - Phys.Rev. C 80, 034323 (2009).}\\
&\parbox[t]{6in}{\addtolength{\parindent}{-0.25cm} \textit{Light nuclei of even mass number in the Skyrme model.}}\\
\href{https://www.nndc.bnl.gov/nsr/nsrlink.jsp?2010Ch42,B}{2010Ch42}&\parbox[t]{6in}{\addtolength{\parindent}{-0.25cm}R.J.Charity, J.M.Elson, J.Manfredi, R.Shane et al. - Phys.Rev. C 82, 041304 (2010).}\\
&\parbox[t]{6in}{\addtolength{\parindent}{-0.25cm} \textit{2p-2p decay of \ensuremath{^{\textnormal{8}}}C and isospin-allowed 2p decay of the isobaric-analog state in \ensuremath{^{\textnormal{8}}}B.}}\\
\href{https://www.nndc.bnl.gov/nsr/nsrlink.jsp?2010Ti04,B}{2010Ti04}&\parbox[t]{6in}{\addtolength{\parindent}{-0.25cm}N.K.Timofeyuk - Phys.Rev. C 81, 064306 (2010).}\\
&\parbox[t]{6in}{\addtolength{\parindent}{-0.25cm} \textit{Overlap functions, spectroscopic factors, and asymptotic normalizationcoefficients generated by a shell-model source term.}}\\
\href{https://www.nndc.bnl.gov/nsr/nsrlink.jsp?2011Ch32,B}{2011Ch32}&\parbox[t]{6in}{\addtolength{\parindent}{-0.25cm}R.J.Charity, J.M.Elson, J.Manfredi, R.Shane et al. - Phys.Rev. C 84, 014320 (2011).}\\
&\parbox[t]{6in}{\addtolength{\parindent}{-0.25cm} \textit{Investigations of three-, four-, and five-particle decay channels of levels in light nuclei created using a \ensuremath{^{\textnormal{9}}}C beam.}}\\
\href{https://www.nndc.bnl.gov/nsr/nsrlink.jsp?2011Ch53,B}{2011Ch53}&\parbox[t]{6in}{\addtolength{\parindent}{-0.25cm}R.J.Charity, J.M.Elson, J.Manfredi, R.Shane et al. - Phys.Rev. C 84, 051308 (2011).}\\
&\parbox[t]{6in}{\addtolength{\parindent}{-0.25cm} \textit{Isobaric multiplet mass equation for A=7 and 8.}}\\
\href{https://www.nndc.bnl.gov/nsr/nsrlink.jsp?2011ChZW,B}{2011ChZW}&\parbox[t]{6in}{\addtolength{\parindent}{-0.25cm}R.J.Charity - Proc.of the 4th Inter.Conf.Proton Emitting Nuclei and Related Topics (PROCON 2011), Bordeaux, France, 6-10 June 2011, LB.Blank Ed. p.87 (2011); AIP Conf.Proc.1409 (2011).}\\
&\parbox[t]{6in}{\addtolength{\parindent}{-0.25cm} \textit{Momentum correlations in the two-proton decay of light nuclei.}}\\
\href{https://www.nndc.bnl.gov/nsr/nsrlink.jsp?2011Pr03,B}{2011Pr03}&\parbox[t]{6in}{\addtolength{\parindent}{-0.25cm}B.Pritychenko, E.Betak, M.A.Kellett, B.Singh, J.Totans - Nucl.Instrum.Methods Phys.Res. A640, 213 (2011).}\\
&\parbox[t]{6in}{\addtolength{\parindent}{-0.25cm} \textit{The Nuclear Science References (NSR) database and Web Retrieval System.}}\\
\href{https://www.nndc.bnl.gov/nsr/nsrlink.jsp?2012My02,B}{2012My02}&\parbox[t]{6in}{\addtolength{\parindent}{-0.25cm}T.Myo, Y.Kikuchi, K.Kato - Phys.Rev. C 85, 034338 (2012); Erratum Phys.Rev. C 87, 049902 (2013).}\\
&\parbox[t]{6in}{\addtolength{\parindent}{-0.25cm} \textit{Five-body resonances of \ensuremath{^{\textnormal{8}}}C using the complex scaling method.}}\\
\href{https://www.nndc.bnl.gov/nsr/nsrlink.jsp?2012My04,B}{2012My04}&\parbox[t]{6in}{\addtolength{\parindent}{-0.25cm}T.Myo - Prog.Theor.Phys.(Kyoto), Suppl. 196, 211 (2012).}\\
&\parbox[t]{6in}{\addtolength{\parindent}{-0.25cm} \textit{Resonances and Continuum States, and Energy Spectra of Light Drip-LineNuclei.}}\\
\href{https://www.nndc.bnl.gov/nsr/nsrlink.jsp?2012Th01,B}{2012Th01}&\parbox[t]{6in}{\addtolength{\parindent}{-0.25cm}M.Thoennessen - At.Data Nucl.Data Tables 98, 43 (2012).}\\
&\parbox[t]{6in}{\addtolength{\parindent}{-0.25cm} \textit{Discovery of isotopes with Z \ensuremath{\leq} 10.}}\\
\href{https://www.nndc.bnl.gov/nsr/nsrlink.jsp?2013La29,B}{2013La29}&\parbox[t]{6in}{\addtolength{\parindent}{-0.25cm}Y.H.Lam, B.Blank, N.A.Smirnova, J.B.Bueb, M.S.Antony - At.Data Nucl.Data Tables 99, 680 (2013).}\\
&\parbox[t]{6in}{\addtolength{\parindent}{-0.25cm} \textit{The isobaric multiplet mass equation for A \ensuremath{\leq} 71 revisited.}}\\
\href{https://www.nndc.bnl.gov/nsr/nsrlink.jsp?2014Eb02,B}{2014Eb02}&\parbox[t]{6in}{\addtolength{\parindent}{-0.25cm}S.Ebata, T.Nakatsukasa, T.Inakura - Phys.Rev. C 90, 024303 (2014); Erratum Phys.Rev. C 92, 069902 (2015).}\\
&\parbox[t]{6in}{\addtolength{\parindent}{-0.25cm} \textit{Systematic investigation of low-lying dipole modes using the canonical-basis time-dependent Hartree-Fock-Bogoliubov theory.}}\\
\href{https://www.nndc.bnl.gov/nsr/nsrlink.jsp?2014Mi17,B}{2014Mi17}&\parbox[t]{6in}{\addtolength{\parindent}{-0.25cm}T.Mizusaki, T.Myo, K.Kato - Prog.Theor.Exp.Phys.\hphantom{a}2014, 091D01 (2014).}\\
&\parbox[t]{6in}{\addtolength{\parindent}{-0.25cm} \textit{A new approach for many-body resonance spectroscopy with the complex scaling method.}}\\
\href{https://www.nndc.bnl.gov/nsr/nsrlink.jsp?2014My03,B}{2014My03}&\parbox[t]{6in}{\addtolength{\parindent}{-0.25cm}T.Myo, K.Kato - Prog.Theor.Exp.Phys.\hphantom{a}2014, 083D01 (2014).}\\
&\parbox[t]{6in}{\addtolength{\parindent}{-0.25cm} \textit{Mirror symmetry breaking in He isotopes and their mirror nuclei.}}\\
\href{https://www.nndc.bnl.gov/nsr/nsrlink.jsp?2017Ka45,B}{2017Ka45}&\parbox[t]{6in}{\addtolength{\parindent}{-0.25cm}K.Kaki - Prog.Theor.Exp.Phys.\hphantom{a}2017, 093D01 (2017).}\\
&\parbox[t]{6in}{\addtolength{\parindent}{-0.25cm} \textit{Reaction cross sections of proton scattering from carbon isotopes (A=8-22) by means of the relativistic impulse approximation.}}\\
\href{https://www.nndc.bnl.gov/nsr/nsrlink.jsp?2017Wa10,B}{2017Wa10}&\parbox[t]{6in}{\addtolength{\parindent}{-0.25cm}M.Wang, G.Audi, F.G.Kondev, W.J.Huang et al. - Chin.Phys.C 41, 030003 (2017).}\\
&\parbox[t]{6in}{\addtolength{\parindent}{-0.25cm} \textit{The AME2016 atomic mass evaluation (II). Tables, graphs and references.}}\\
\href{https://www.nndc.bnl.gov/nsr/nsrlink.jsp?2019Ka50,B}{2019Ka50}&\parbox[t]{6in}{\addtolength{\parindent}{-0.25cm}C.Karthika, M.Balasubramaniam - Phys.Rev. C 100, 054611 (2019).}\\
&\parbox[t]{6in}{\addtolength{\parindent}{-0.25cm} \textit{Mirror nuclei of 1n/2n halo systems as 1p/2p emitters.}}\\
\href{https://www.nndc.bnl.gov/nsr/nsrlink.jsp?2019Sh36,B}{2019Sh36}&\parbox[t]{6in}{\addtolength{\parindent}{-0.25cm}P.G.Sharov, L.V.Grigorenko, A.N.Ismailova, M.V.Zhukov - JETP Lett. 110, 5 (2019).}\\
&\parbox[t]{6in}{\addtolength{\parindent}{-0.25cm} \textit{Pauli-Principle Driven Correlations in Four-Neutron Nuclear Decays.}}\\
\href{https://www.nndc.bnl.gov/nsr/nsrlink.jsp?2021My01,B}{2021My01}&\parbox[t]{6in}{\addtolength{\parindent}{-0.25cm}T.Myo, M.Odsuren, K.Kato - Phys.Rev. C 104, 044306 (2021).}\\
&\parbox[t]{6in}{\addtolength{\parindent}{-0.25cm} \textit{Five-body resonances in \ensuremath{^{\textnormal{8}}}He and \ensuremath{^{\textnormal{8}}}C using the complex scaling method.}}\\
\href{https://www.nndc.bnl.gov/nsr/nsrlink.jsp?2021Wa16,B}{2021Wa16}&\parbox[t]{6in}{\addtolength{\parindent}{-0.25cm}M.Wang, W.J.Huang, F.G.Kondev, G.Audi, S.Naimi - Chin.Phys.C 45, 030003 (2021).}\\
&\parbox[t]{6in}{\addtolength{\parindent}{-0.25cm} \textit{The AME 2020 atomic mass evaluation (II). Tables, graphs and references.}}\\
\href{https://www.nndc.bnl.gov/nsr/nsrlink.jsp?2021Wy01,B}{2021Wy01}&\parbox[t]{6in}{\addtolength{\parindent}{-0.25cm}J.Wylie, J.Okolowicz, W.Nazarewicz, M.Ploszajczak et al. - Phys.Rev. C 104, L061301 (2021).}\\
&\parbox[t]{6in}{\addtolength{\parindent}{-0.25cm} \textit{Spectroscopic factors in dripline nuclei.}}\\
\href{https://www.nndc.bnl.gov/nsr/nsrlink.jsp?2021Xi06,B}{2021Xi06}&\parbox[t]{6in}{\addtolength{\parindent}{-0.25cm}F.Xing, J.Cui, Y.Wang, J.Gu - Chin.Phys.C 45, 124105 (2021).}\\
&\parbox[t]{6in}{\addtolength{\parindent}{-0.25cm} \textit{Two-proton radioactivity of ground and excited states within a unifiedfission model.}}\\
\href{https://www.nndc.bnl.gov/nsr/nsrlink.jsp?2022De06,B}{2022De06}&\parbox[t]{6in}{\addtolength{\parindent}{-0.25cm}D.S.Delion, S.A.Ghinescu - Phys.Rev. C 105, L031301 (2022).}\\
&\parbox[t]{6in}{\addtolength{\parindent}{-0.25cm} \textit{Two-proton emission systematics.}}\\
\href{https://www.nndc.bnl.gov/nsr/nsrlink.jsp?2022Zo01,B}{2022Zo01}&\parbox[t]{6in}{\addtolength{\parindent}{-0.25cm}Y.Y.Zong, C.Ma, M.Q.Lin, Y.M.Zhao - Phys.Rev. C 105, 034321 (2022).}\\
&\parbox[t]{6in}{\addtolength{\parindent}{-0.25cm} \textit{Mass relations of mirror nuclei for both bound and unbound systems.}}\\
\href{https://www.nndc.bnl.gov/nsr/nsrlink.jsp?2023Ch22,B}{2023Ch22}&\parbox[t]{6in}{\addtolength{\parindent}{-0.25cm}R.J.Charity, K.Brown, T.Webb, L.G.Sobotka - Phys.Rev. C 107, 054301 (2023).}\\
&\parbox[t]{6in}{\addtolength{\parindent}{-0.25cm} \textit{Invariant-mass spectroscopy of \ensuremath{^{\textnormal{10}}}B, \ensuremath{^{\textnormal{11}}}C, \ensuremath{^{\textnormal{14}}}F, \ensuremath{^{\textnormal{16}}}F, and \ensuremath{^{\textnormal{18}}}Na.}}\\
\href{https://www.nndc.bnl.gov/nsr/nsrlink.jsp?2023Ch46,B}{2023Ch46}&\parbox[t]{6in}{\addtolength{\parindent}{-0.25cm}R.J.Charity, J.Wylie, S.M.Wang, T.B.Webb et al. - Phys.Rev.Lett. 131, 172501 (2023).}\\
&\parbox[t]{6in}{\addtolength{\parindent}{-0.25cm} \textit{Strong Evidence for \ensuremath{^{\textnormal{9}}}N and the Limits of Existence of Atomic Nuclei.}}\\
\href{https://www.nndc.bnl.gov/nsr/nsrlink.jsp?2023My01,B}{2023My01}&\parbox[t]{6in}{\addtolength{\parindent}{-0.25cm}T.Myo, K.Kato - Phys.Rev. C 107, 014301 (2023).}\\
&\parbox[t]{6in}{\addtolength{\parindent}{-0.25cm} \textit{Possible interpretation of the complex expectation values associated with resonances.}}\\
\href{https://www.nndc.bnl.gov/nsr/nsrlink.jsp?2023My02,B}{2023My02}&\parbox[t]{6in}{\addtolength{\parindent}{-0.25cm}T.Myo, K.Kato - Phys.Rev. C 107, 034305 (2023).}\\
&\parbox[t]{6in}{\addtolength{\parindent}{-0.25cm} \textit{Soft dipole resonance in \ensuremath{^{\textnormal{8}}}C and its isospin symmetry with \ensuremath{^{\textnormal{8}}}He.}}\\
\href{https://www.nndc.bnl.gov/nsr/nsrlink.jsp?2024Ko04,B}{2024Ko04}&\parbox[t]{6in}{\addtolength{\parindent}{-0.25cm}S.Koyama, D.Suzuki, M.Assie, L.Lalanne et al. - Phys.Rev. C 109, L031301 (2024).}\\
&\parbox[t]{6in}{\addtolength{\parindent}{-0.25cm} \textit{Mirror symmetry at far edges of stability: The cases of \ensuremath{^{\textnormal{8}}}C and \ensuremath{^{\textnormal{8}}}He.}}\\
\href{https://www.nndc.bnl.gov/nsr/nsrlink.jsp?2024Ya25,B}{2024Ya25}&\parbox[t]{6in}{\addtolength{\parindent}{-0.25cm}T.Yarman, O.Yarman, N.Zaim, A.Kholmetskii, M.Arik - Int.J.Mod.Phys. E33, 2450032 (2024).}\\
&\parbox[t]{6in}{\addtolength{\parindent}{-0.25cm} \textit{Systematization of \ensuremath{\beta}\ensuremath{^{\textnormal{+}}}-decaying atomic nuclei: Interrelation between half-life, mass, energy and size.}}\\
\end{longtable}
\end{document}